\documentclass[structabstract]{aa}  
\usepackage{graphicx}
\usepackage{savesym}
\usepackage{amsmath}
\usepackage{amssymb}
\usepackage{txfonts}
\usepackage{time}
\usepackage[usenames]{color}
\usepackage{xspace}

\usepackage{natbib}
\bibpunct{(}{)}{;}{a}{}{,}

\def\setid#1,v #2 #3/#4/#5 #6 #7 #8.{(\textit{CVS version #2, checked in by #7 on #3-#4-#5, #6 UT})}

\newcommand{\MURaM}{\texttt{MURaM}\xspace}
\newcommand{\kms}{~km\,s$^{-1}$}

\newcommand{\sref}[1]{Sec.~\ref{#1}}
\newcommand{\fig}[1]{Fig.~\ref{#1}}

\newcommand{\caiih}{\ion{Ca}{ii}\,H}
\newcommand{\FeI}{\ion{Fe}{i}}

\newcommand{\halpha}{\ion{H$\alpha$}{}}
\def\tlc{^{10}\log\tau_c}

\newcommand{\corrone}[1]{{#1}}
\newcommand{\corrtwo}[1]{{#1}}

\begin{document}
\bibliographystyle{aa}
\frenchspacing
\title{Peripheral downflows in sunspot penumbrae}
\titlerunning{Peripheral downflows}
 
\author{M.\ van Noort\inst{1} \and A. Lagg\inst{1} \and S. K. Tiwari\inst{1} \and S. K. Solanki\inst{1,2}}

\date{Draft: \now\ \today\ \setid $Id: ms.tex,v 1.103 2013/07/17 09:23:24 noort Exp $.}
                  
\offprints{M. v. N.: \email{vannoort@mps.mpg.de}}

\institute{Max Planck Institute for Solar System Research, Max-Planck Stra\ss e 2, D-37191 Katlenburg-Lindau, Germany\\
           \and
           School of Space Research, Kyung Hee University, Yongin, Gyeonggi 446-701, Republic of Korea}

\abstract
{Sunspot penumbrae show high-velocity patches along the periphery.}
{The high-velocity downflow patches are believed to be the return channels of the Evershed flow. We aim to investigate their structure in detail using Hinode SOT/SP observations.}
{We employ Fourier interpolation in combination with spatially coupled height dependent LTE inversions of Stokes profiles to produce high-resolution, height-dependent maps of atmospheric parameters of these downflows and investigate their properties.}
{High-speed downflows are observed over a wide range of viewing angles. They have supersonic line-of-sight velocities, some in excess of 20\kms{}, and very high magnetic field strengths, reaching {\corrone{values of over}} 7 kG. A relation between the downflow velocities and the magnetic field strength is found, in good agreement with MHD simulations.}
{The coupled inversion at high resolution allows for the accurate determination of small-scale structures. The recovered atmospheric structure indicates that regions with very high downflow velocities contain some of the strongest magnetic fields that have ever been measured on the Sun.}


\maketitle

\section{Introduction}
Since the discovery of the Evershed effect \citep{ever09}, the origin and mass flow budget of sunspot penumbrae have been the topic of much debate. In particular, the location of the return flow of the material to the photosphere was extensively debated, owing to a shortage of conclusive observational evidence \citep[see][for an overview]{sola03}. 

There has been progress with the discovery of compact downflow regions in the outer parts of the penumbra \citep{west97,west01,trit04,mart09,fran09}, suggesting that the flow returns there to the deep photosphere. It was shown by \cite{sola94}, using Stokes $I$ and $V$ profiles of the 1.56~$\mu$m line, that the majority of the material carried by the Evershed flow returns below the surface within the penumbra and only a small fraction continues into the canopy.

However, the very compact nature of these flows has made it challenging to studying them in detail. The observed data suffer from degradations, which are caused by the limited spatial resolution of the telescope that was used to record them, the effect of which is mixing the spectra of neighboring pixels, which complicates their interpretation. Consequently, the observed maximum flow velocities have increased, as has the spatial resolution of the data \citep{ichi07a}. This trend continues to this day, since even the highest resolution data cannot claim to fully resolve these downflows \citep{jurc10}. 



Using data from the Advanced Stokes Polarimeter (ASP), \cite{vmp94} observed downflows as large as 14\kms in the umbra of a delta sunspot, which they took to be the signature of supersonic compressive fluid flows in the lower to mid photosphere. These flows were attributed to the draining and subsequent funneling of material from a rising magnetic loop system. \cite{mart09} find that mostly horizontal field channels just outside sunspot penumbrae sustain supersonic flows with a line-of-sight component of up to 6\kms{} and interpret these flows as the continuation of the Evershed flow outside the spot. \cite{wiehr96} suggested that the dark, compact downflow regions outside sunspot penumbrae are the result of a sudden change in the equipartition between kinetic and magnetic energy densities.

The observational signatures of penumbral downflows have also been studied using theoretical approaches. \cite{thom02} report on strong, coherent, descending plumes resulting from turbulent, compressible convection at the outer edge of a simulated sunspot. In the most recent magneto-hydrodynamic (MHD) simulations \citep{remp12}, the average velocity in the supersonic downflow regions is found to be 9.6\kms{} at optical depth unity, with the fastest flows in the outer regions of the penumbra reaching up to 15\kms. However, the contribution of these fast downflows to the total downward directed mass flux in the penumbra is found to be negligible.

\begin{figure*}[ht]
\includegraphics[width=\textwidth]{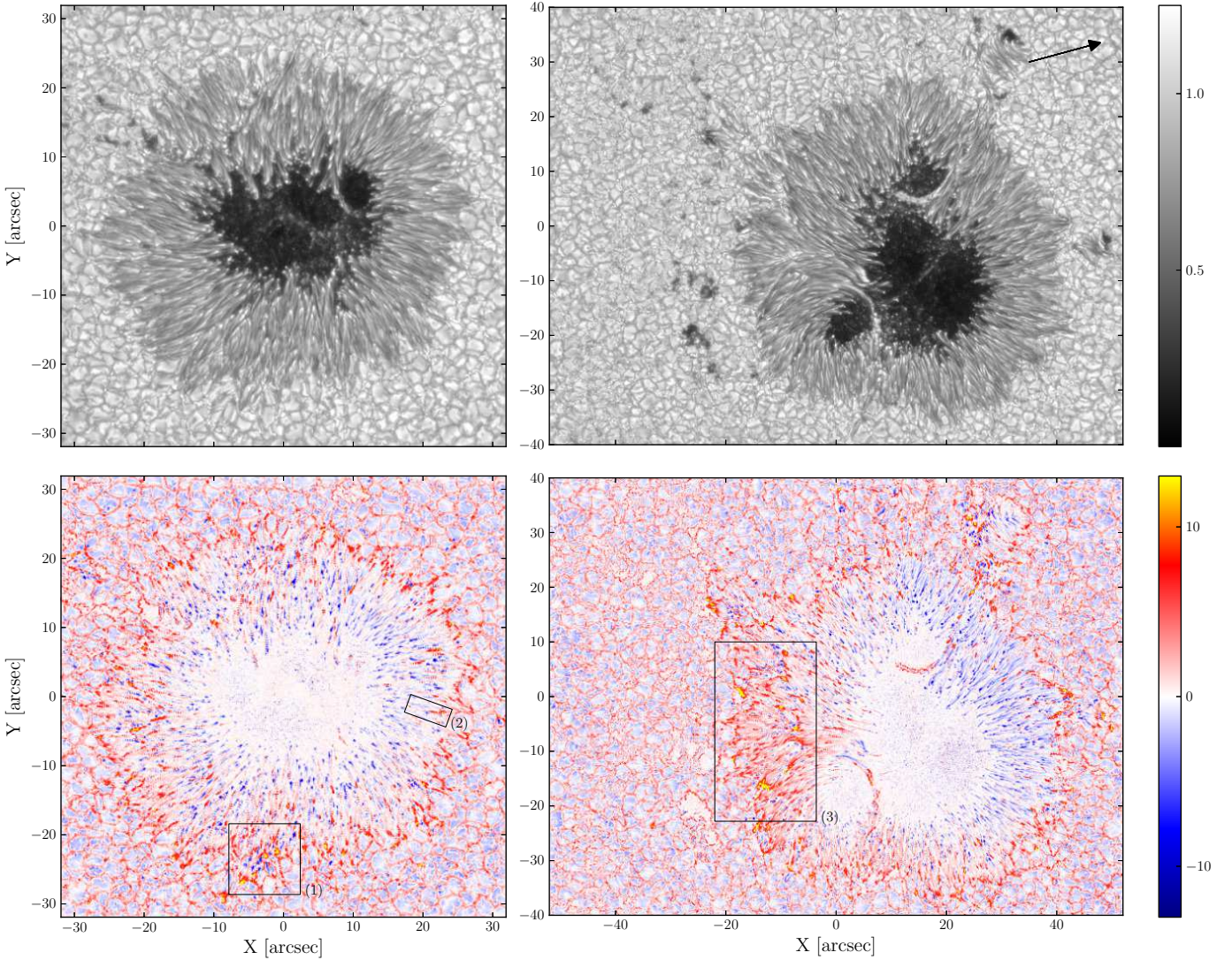}
\caption{Continuum intensity map (top) and LOS velocity at continuum optical
  depth unity (bottom) for spot 1 (left) and spot 2 (right). The boxes
  indicate the selected regions of interest discussed in the text. The arrow points in the direction of disk center for spot 2. Spot 1 is observed almost at disc center.}
\label{fig:overview}
\end{figure*}

Although most reports of downflows find them to contain magnetic fields with a polarity opposite that of the magnetic field in the umbra \citep{west97,ichi07a,borr11}, a new type of downflow has recently been reported to contain a magnetic field with the same polarity as the magnetic field in the umbra \citep{kats10}. These downflows reach maximum velocities of up to 8\kms{} in 1.6--6~arcsec$^2$ sized patches in the inner penumbra of a sunspot and have also been reported by \cite{loui12}.

\cite{fran09} determined downflow velocities up to 9\kms{} in the leading spot of NOAA AR 10933, the same as one of the sunspots, which we investigate in the present paper. They used line bisectors of the wings of the absorption lines to assess the flow patterns. Dominating upflow and downflow patches observed in the inner and outer penumbra, respectively, were interpreted as the sources and sinks of the Evershed flow.

In this paper, we re-examine the data used by the above authors, but use a spatially coupled inversion technique \citep[][paper I]{vannoort12}, so that the spectral contamination introduced by the spatial degradation caused by the telescope is properly taken into account. We also apply this technique to another sunspot, the trailing spot of NOAA AR 10953 and compare the observed downflows with the recent MHD simulations by \cite{remp12}.

\section{Observations}
\label{sec:obs}

For this study we use datasets obtained with the spectropolarimeter \citep{lite01} of the Hinode Solar Optical Telescope \citep[SOT/SP,][]{tsun08,kosu07}. The data are scanned slit-spectra of the Fe I lines at 6301.5 and 6302.5\,\AA, with a typical signal-to-noise level of 1000 and a spatial \corrone{sampling} of 0.16''.

We investigate the leading spot of NOAA AR 10933 (spot 1), observed only 2 degrees off disc center on January 05$^{th}$ 2007 from 1236--1310 UT, and the trailing spot of NOAA AR 10953 (spot 2), observed at position ($-342$\arcsec{},$-98$\arcsec{}, $\mu=\cos\theta=0.92$) on April 30$^{th}$ 2007 from 1835--1939 UT (see \fig{fig:overview}). 

The proximity to disc center of spot 1 aids in the interpretation of the data and allows us to analyze features from all across the sunspot without being hindered by projection effects, while spot 2 allows us to study any possible effects that require a moderately inclined viewing angle. Spot 1 has a positive magnetic polarity, whereas that of spot 2 is negative. The SP data were processed using the standard Hinode reduction tools from SolarSoft.

\section{Inversions}
\label{sec:inv}
The data were inverted using the SPINOR inversion code \citep{2000PhDTh,frut00}, based on the STOPRO routines \citep{sola87}, in the spatially coupled mode described in paper I. This technique takes the telescope diffraction directly into account in the inversion process. Instead of fitting the observed Stokes profiles for every individual pixel of the map, the synthetic profiles are computed over a two-dimensional subfield, then convolved with the telescope point spread function (PSF), and finally matched to the observed Stokes profiles of the whole subfield simultaneously. This technique successfully reproduces complex Stokes profiles with a simple, three node atmospheric model where conventional, pixel-based inversion techniques require multi-component atmospheres to achieve a fit to the observed profiles of similar quality.

The PSF, required for the spatially coupled inversion of the data, was constructed from the Hinode pupil function \citep{suem08}, with 0.1 waves defocus added to account for the observed quiet sun contrast \citep{dani08}. The atmospheric temperature $T$, line-of-sight velocity $v_{\mbox{LOS}}$, magnetic field strength $B$, inclination $\gamma$ and azimuth $\phi$, and a micro-turbulent velocity $v_{\mbox{mic}}$ were fitted at three height nodes located at $\tlc=(-2.5,-0.9,0.0)$ for spot 1 and $\tlc=(-2.5,-0.8,0.0)$ for spot 2, where $\tau_c$ is the optical depth of the local continuum at a wavelength of 6302.5\AA. No macro-turbulent velocity was fitted.

\corrone{Although the placement of the optical depth nodes is fairly flexible in the SPINOR inversion code, the number of nodes must be either a single one, indicating constant properties as a function of height, or otherwise at the same optical depth values for all height-dependent fitted quantities. Although the spectral information would allow for several height nodes in temperature, it does not support more than three nodes in the magnetic field parameters. Therefore, the number of nodes was limited to three, making the interpolated temperature stratification relatively sensitive to the placement of the central node. Best fits of the spectra were found with a placement of the middle node at -0.9 for spot 1, located at disc center, and slightly deeper at -0.8 for spot 2 at $\mu=0.92$.}

Maps of the inverted continuum intensity and velocity at continuum optical depth unity of spot 1 and spot 2 are shown in \fig{fig:overview}. Numerous strong, positive LOS velocity patches can be seen around the interface between the penumbra and the surrounding moat, which, by the convention used here, correspond to downflows. Spot 2 shows a similar pattern to spot 1, but with many particularly strong and large downflows, concentrated on the limb side of the sunspot. In addition, in contrast to spot 1, in this spot the Evershed flow is clearly visible.

\begin{figure}[htb]
\includegraphics[width=9.00cm]{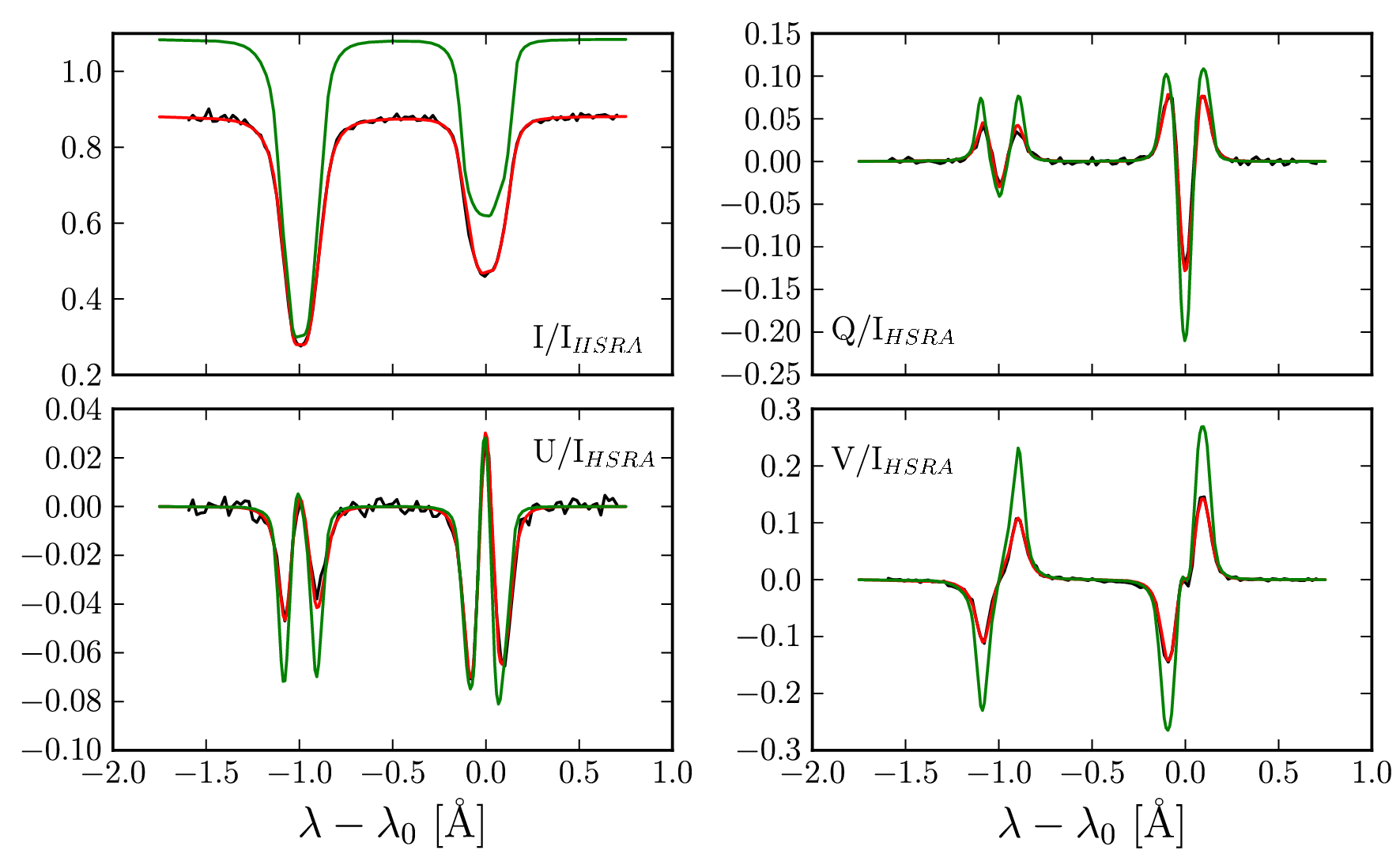}
\caption{Observed (black), fitted (green) and convolved fitted (red) profiles at a ``typical'' location $[x,y]=[-4\,\farcs{}72,-9\,\farcs{}52]$ in box(3) in \fig{fig:overview}, i.e. a location displaying a single atmospheric component and only moderate flow velocity. Clearly, the convolved fit and the data match very well, leaving a large discrepancy between the actual profile (green) and the observed one.}
\label{fig:profiles2}
\end{figure}

A typical profile from the penumbral region of spot 2, fitted using the coupled inversion technique, is plotted in \fig{fig:profiles2}. The observed profiles are reproduced nearly perfectly by the convolved profiles, but show substantial differences with the actually fitted, unconvolved profiles in that location. The magnitude of this discrepancy is also typical and is reduced only in the central parts of large granules, where, due to the uniformity of the Stokes profiles, the effect of the convolution is relatively small.

\begin{figure*}[htb]
\includegraphics[width=18.00cm]{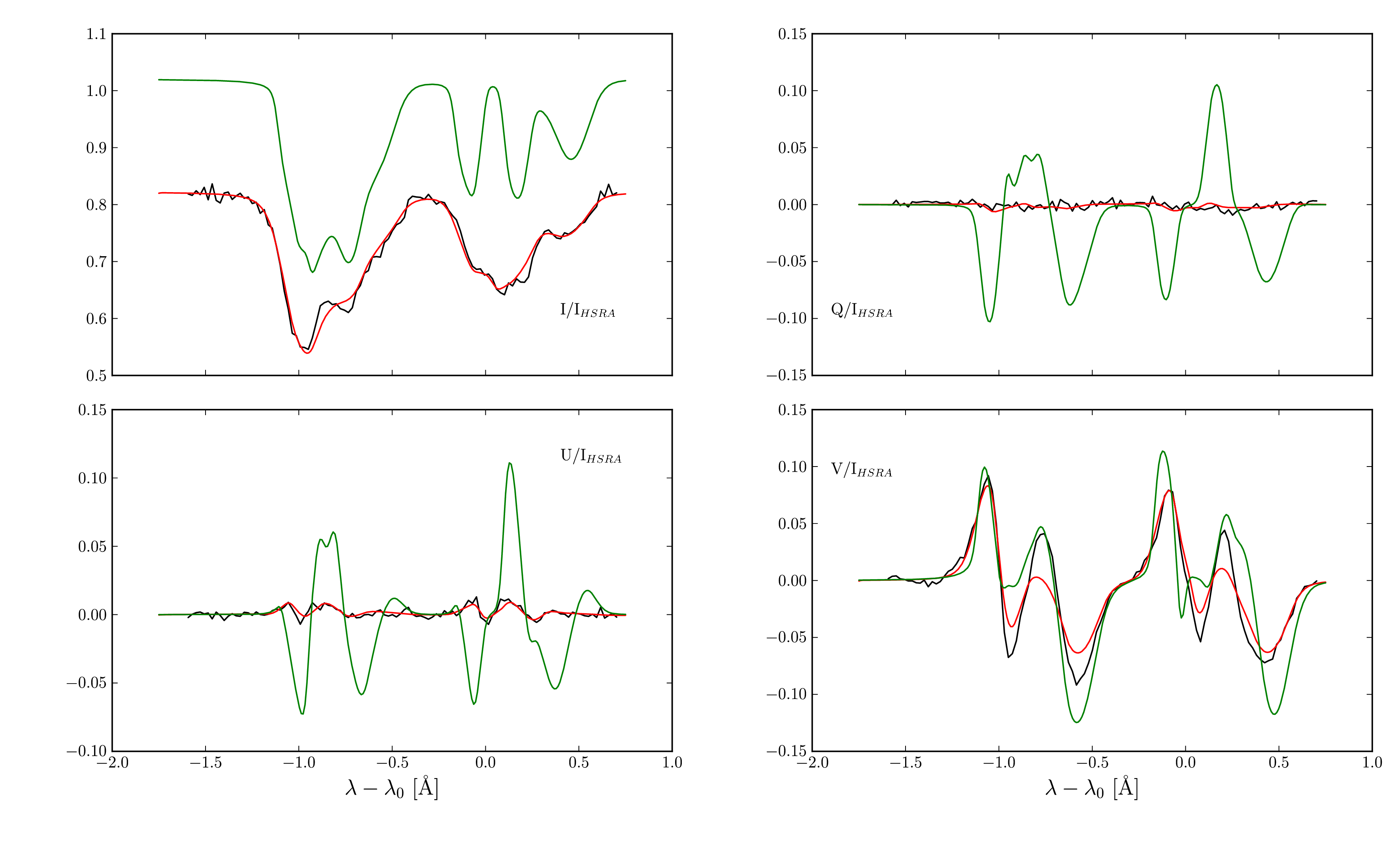}
\caption{Observed (black), fitted (green) and convolved fitted (red) profiles at an ``untypical'' location $[x,y]=[-13\,\farcs{}36,-17\,\farcs{}28]$ in box (3) in \fig{fig:overview} where the observed profiles show clear signs of multiple components. Clearly the coupled inversion fit captures the basic behavior of the profile well. Also note the very large differences between the convolved profiles and the actual ones. \corrone{The locally fit values at optical depth unity are $T=6.4$~kK, $B=7.5$~kG, $\gamma=30^\circ$ and $v_{los}=10.9\,km/s$.}}
\label{fig:profiles1}
\end{figure*}

The profiles of the downflows we are interested in here, such as the one shown in \fig{fig:profiles1}, are reminiscent of those produced by a multi-component atmosphere. Although all fitted atmospheres in the coupled inversion are single component atmospheres and the contribution of each pixel to the spectra of the surrounding pixels is completely prescribed by the telescope PSF, the data are fitted very well, although arguably not as accurately as for the majority of the ``typical'' pixels. The remaining discrepancy between the data and the fitted profile may be attributed to any of a number of causes, such as unresolved substructure in the vertical or the horizontal direction. Despite this, the fitted profiles show the qualitatively correct behavior and are remarkably accurate considering that only single-component, three-node atmospheres were used.

The most extreme looking profiles, such as the example profiles from \fig{fig:profiles1}, can be found in spot 2, in downflow areas on the limb side of the spot. These downflows are among the largest and strongest downflow areas found in both the active regions and contain profiles that show the Fe I lines at 6301.5 and 6302.5~\AA{} depressed to such an extent that there is no clearly discernible continuum left between the two lines. Even these profiles are fitted fairly well by the inversion, without any change in the atmospheric model, although the atmospheric model may need to be adjusted to reach the quality of fit that is achieved almost everywhere else (which has not been attempted for the inversions here).

\begin{figure*}[htb]
\includegraphics[width=\textwidth]{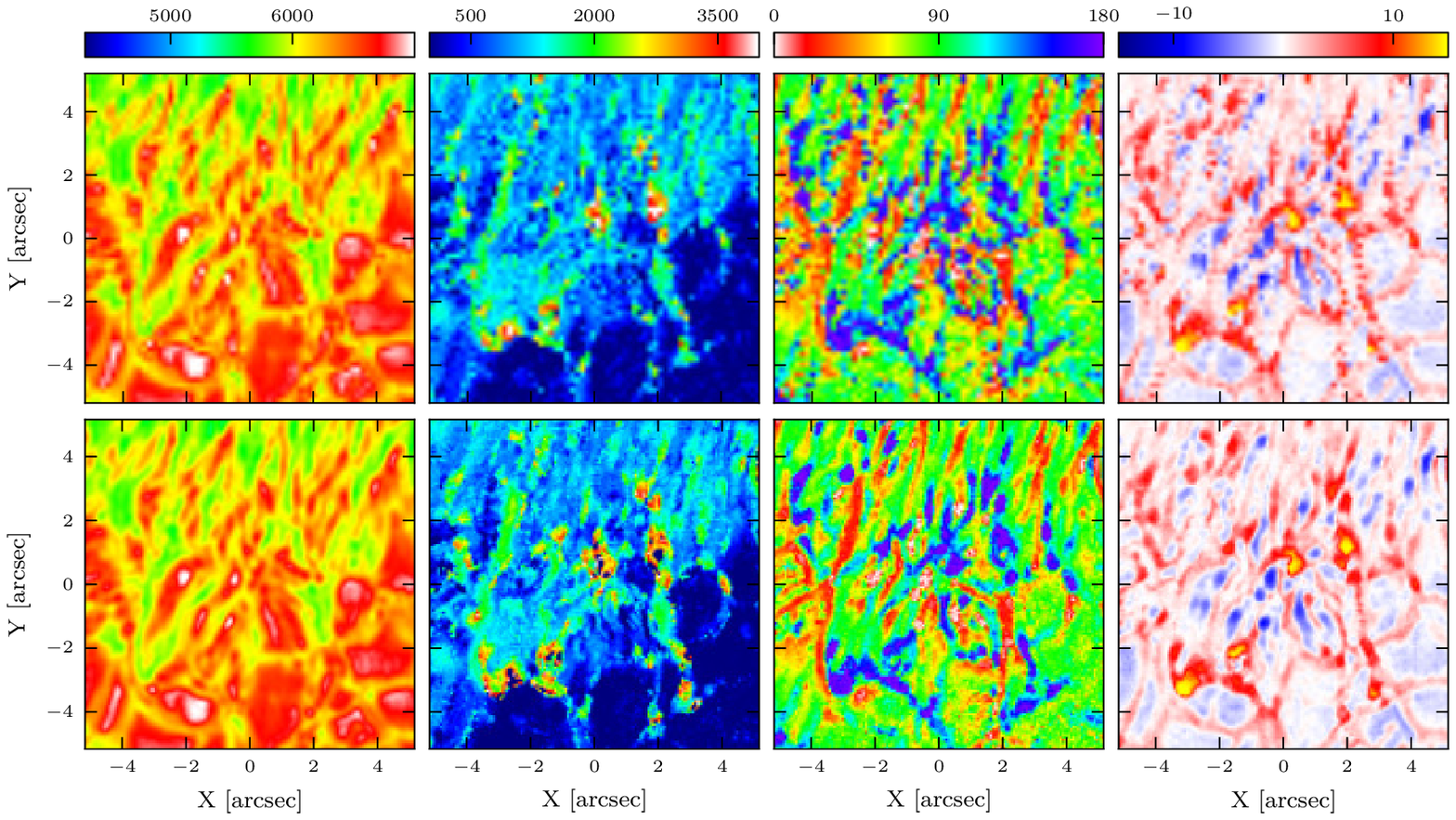}
\caption{Selected region of spot 1, indicated by a black square in \fig{fig:overview}, at the native \corrone{sampling} of Hinode SP (top) and interpolated to 0{\,\farcs}08$/$pixels prior to inversion (bottom). From left to right: temperature [K], magnetic field strength [G], inclination [$^\circ$], and line-of-sight velocity [\kms{}], all at $\tlc=0$. Clearly visible is the increased coherence of the structures and reduced overall noise in almost all parameters retrieved from the interpolated data.}
\label{fig:blowup}
\end{figure*}

Despite the high quality of the fits, the velocity structure in the inverted maps shows significant oscillatory behavior on the scale of a pixel, as can be seen in \fig{fig:blowup}\corrone{, for example around location $(x,y)=(2,-1)$ in the line-of-sight velocity}. In this blow-up of spot 1 (region (1) in \fig{fig:overview}) at continuum optical depth unity, numerous compact, supersonic downflows with LOS velocities of up to 18\kms{} are present. Although the large-scale behavior seems realistic, many small-scale fluctuations show spatial oscillations or appear quite noisy (see also paper I for a description of this problem), despite giving an excellent fit to the data. Additionally, most of the spectra in these regions show complicated profiles, sometimes with clearly separable components, as if they are produced by separate structures in the atmosphere.

The appearance of localized but persistent oscillations and physically implausible structures, in combination with the multi-component character of the spectra suggests that we may be dealing with a substantial amount of substructure in the atmosphere, which is clearly visible in the spectra, but which cannot be resolved on the pixel grid defined by the observations.

\subsection{Hinode at 0{\,\farcs}08}
As suggested in the previous section, there are indications that the inverted atmosphere may be lacking a significant amount of substructure that cannot be resolved on the pixel grid of the observed data, but that nonetheless manifests itself in the observed spectra. The question thus presents itself if this situation can be improved upon by artificially refining the pixel grid of the solution to represent some of this substructure. This approach resembles inverting the entire dataset using a multi-component atmosphere, but with the additional restrictions that the filling factor of each ``component'' is fixed and that the location of each ``component'' is different from all the others. In addition, the same component contributes to many different pixels via the PSF, further reducing the degree of freedom of the solution. The additional restrictions confine each atmospheric ``component'' to the location that best matches the spatial distribution of the corresponding spectral feature in the observed data. 

As outlined briefly in paper I, the spatially coupled inversion method is stable to oversampled data and produces an inversion result with a resolution up to the resolution limit of the telescope, indicating that the inversion remains sufficiently well constrained in this situation. The problem then remains how to obtain a sufficiently densely sampled dataset, since no appropriately sampled Hinode data are available. To overcome this problem, a Fourier interpolation method was used that is able to produce a dataset at an arbitrary resolution with the appropriate noise characteristics from critically sampled data.

\subsubsection{Interpolation}
\begin{figure}[htb]
\includegraphics[width=9.00cm]{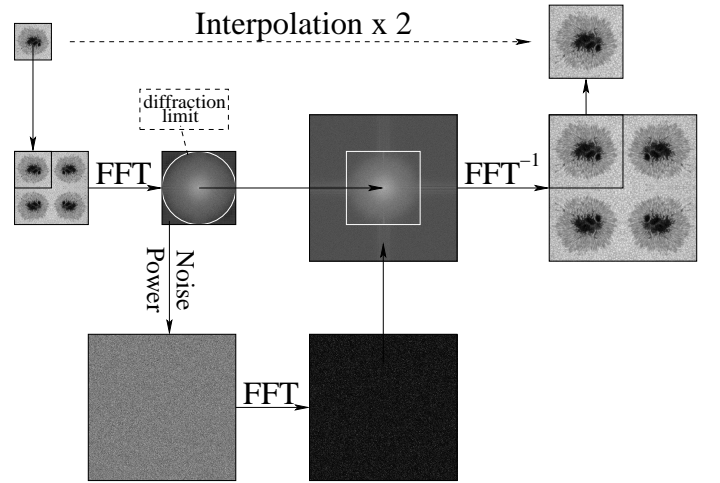}
\caption{Interpolation scheme for Hinode SP data to 0{\,\farcs}08$/$pixel. The scheme illustrated here is for a scale factor \corrone{of N=2}, but in principle any scale change is possible in this way.}
\label{fig:interpol}
\end{figure}
To arrive at a dataset that is as close as possible to the one that would have been obtained if a camera with a smaller pixel size had been used, the procedure illustrated in \fig{fig:interpol} was used. To make the target FOV periodic, it is enlarged by a factor 2 in all directions, by appending an appropriately mirrored duplicate of the data. The thus obtained periodic data cube is then Fourier transformed, which can now be done without generating severe artifacts, since the cube is actually periodic, after which the noise power is calculated from the area of the power spectrum outside the diffraction limit (the white ring in \fig{fig:interpol}).

A data cube containing white noise with a power level equal to that measured in the data cube is then created with an extent $N$ times the size of the periodic data cube (so 2$N$ times the original data cube) and Fourier transformed. The low-frequency data values are then replaced by those from the Fourier transformed periodic data cube, after which the resulting data cube is transformed back. One quarter of the resulting scaled data cube now contains the original data, but at $N$ times the resolution and with a noise level consistent with detection at an identical illumination level but with $1/N$th the pixel size of the original detector.

The addition of the noise is essential, as the omission of this noise creates a pattern in the data at the characteristic spatial scale of the original pixels. The inversion will contain this pattern, but since it is now spatially resolved, it does not look like noise but like statistically significant features.

\subsubsection{Inversions at 0{\,\farcs}08}

Due to the scaling properties of the coupled Levenberg-Marquardt inversion step, the execution time is significantly larger than for the \corrtwo{natively sampled} problem above, with the execution time needed to calculate the approximate operator ${\bf A}^*$  from paper I increasing approximately 16 fold for a factor of 2 reduction in pixel size.

In return, however, the resulting inverted atmosphere shows significantly more detail than was the case at the original resolution and has virtually no oscillatory problems, as can be seen in \fig{fig:blowup}, where several selected inverted parameter maps are shown for the native resolution and double resolution data.
\begin{figure*}[htb]
\includegraphics[width=\textwidth]{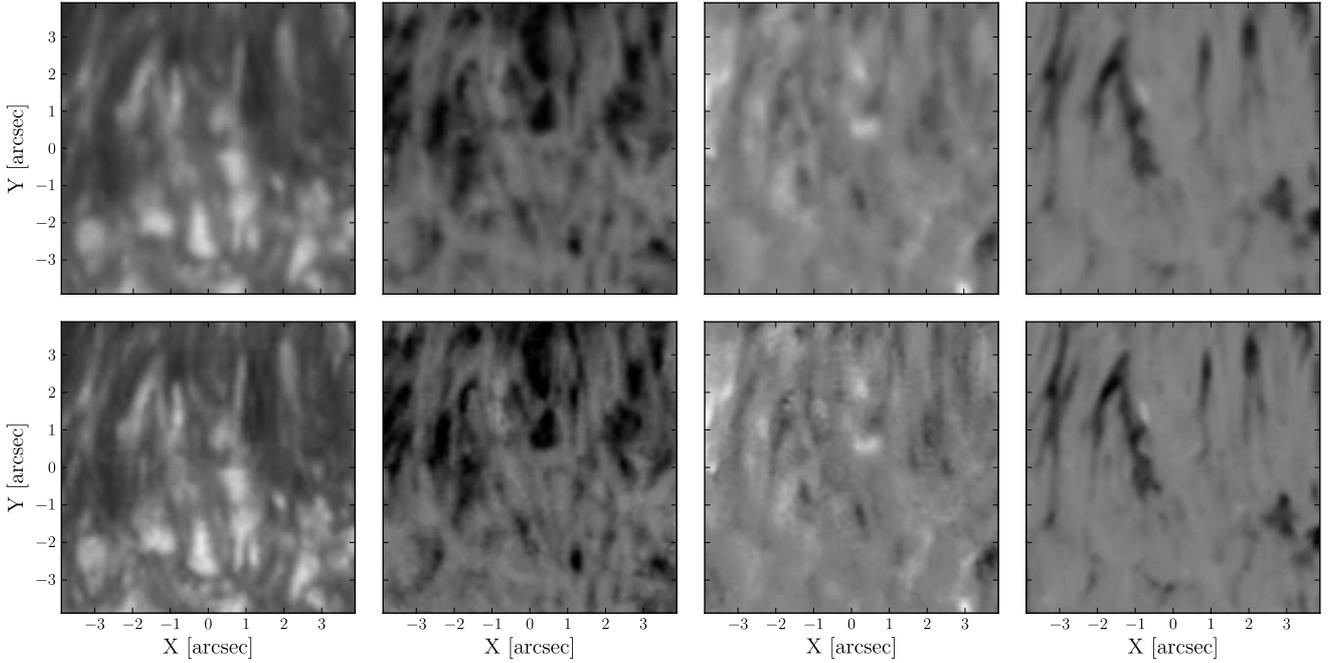}
\caption{Synthetic Stokes images (from left to right: $I$, $Q$, $U$, $V$) in the red wing of the \FeI{} 6301.5~\AA\ line generated from the inversion results at native \corrone{sampling} of the dataset (top) and at 0.08''/pixel (bottom). The gray scale of the images is [0.1,1.1] for $I$, [$-$0.05,0.05] for $Q$ and $U$ and [$-$0.25,0.25] for $V$. The denser \corrone{sampled} image has a visibly higher contrast and contains finer details than the natively \corrone{sampled} image.}
\label{fig:hinode2dhr}
\end{figure*}
\begin{figure*}[htb]
\includegraphics[width=\textwidth]{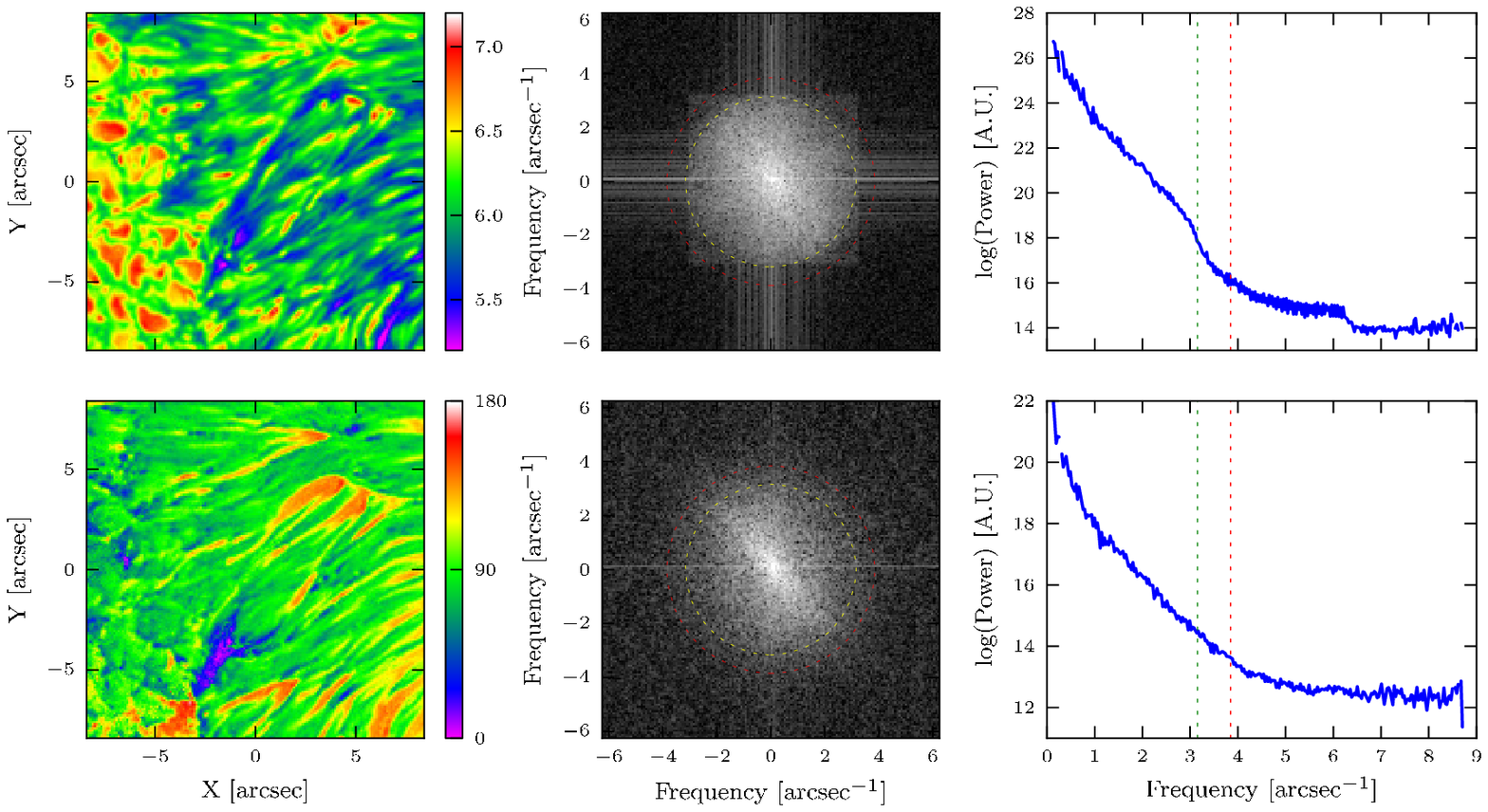}
\caption{High-resolution maps of temperature at $\tau=1$ (top) and inclination at $\tau\approx 0.15$ (bottom), the 2D power spectra (middle) and the azimuthally averaged power spectra of the maps (right). \corrone{The yellow dashed circle (middle panel) and the green (right panel)} and red dashed lines indicate the Rayleigh diffraction limit and $\lambda/D$ respectively.}
\label{fig:power}
\end{figure*}

\fig{fig:hinode2dhr} shows Stokes images calculated from the inverted results at native and double \corrtwo{sampling} in the red wing of the \FeI{} 6301.5~\AA\ line. Although the images both look smooth due to the bicubic spline interpolation used for displaying, the \corrtwo{oversampled} images clearly contain more contrast and detail than the \corrtwo{natively sampled} ones, with some of the more prominent small-scale features clearly reduced in size. Without interpolating the images, this difference is even more striking, due to pixelation effects.

There is a significant difference in the way in which the various fit parameters respond to the increased resolution. Those parameters that do not depend much on the shape of the spectral lines, such as the temperature in the deepest layers, show only a limited amount of power at the frequencies not represented by the original data, as shown in the top row of \fig{fig:power}. The power in the square covered by the original data is clearly visible in the 2D power spectrum, but a noticeable absence of power is observed everywhere else. In the azimuthally averaged power spectrum, this absence of power produces a smooth but noticeable decrease of power between the Rayleigh diffraction limit at 0{\,\farcs}32 (3.1 arcsec$^{-1}$) and the true telescope diffraction limit at 0{\,\farcs}26 (3.8 arcsec$^{-1}$). Other parameters, such as the inclination, shown in the bottom panel of \fig{fig:power}, that are strongly determined by the shape of the spectral lines, have a power spectrum that extends smoothly across the resolution limit of the original data up to and even slightly beyond the telescope diffraction limit. Most of the inverted quantities show only a small amount of power beyond the telescope diffraction limit before disappearing in the noise, suggesting that not much information was retrieved there.

Apart from the interpolation, the same procedures were used to obtain the results in the \corrone{natively sampled and oversampled} inversions. The strong reduction of the noise patterns in the double resolution compared to the native resolution inversion suggests that accurate placement of spectral features in the FOV is important for producing a horizontally coherent atmospheric structure.

Interestingly, the fit to the data does not always look improved and in a few pixels actually appears to be worse than before. 
Although this appears not to speak in favor of this approach, it is in fact a logical consequence of the global optimization strategy used for the spatially coupled inversions. Not the fit of each individual profile, but the simultaneous fit of all profiles in the FOV is optimized, sometimes at the expense of the fit to a few isolated profiles.

On average, the fit of the inverted profiles to the interpolated data is equally good or only slightly improved over the native resolution, however, the consistency of the high-resolution inversion result is much improved, as can be seen by direct comparison of the two solutions in \fig{fig:blowup}: The high-resolution solution shows a lot more detail. The main influence of the added resolution appears to be the increased ability of the inversion code to produce a spectral feature in the correct location rather than of a reduced spatial extent, resulting in spatially coherent rather than apparently random or oscillatory structures.

\corrone{The issue of establishing the error in the fitted atmospheric parameters is a difficult one. Although it is possible to calculate the formal error in each quantity independently, the number of dependencies in a spatially coupled inversion is many times larger than in a one-dimensional inversion. In particular, the modification of one parameter in one pixel may well lead to change in the same parameter in the neighboring pixel due to the spatial coupling. To fully explore this issue would require the solution to be computed for a perturbation of each parameter in every pixel, which is well beyond the reach of our computational resources. Moreover, the resulting formal error would be no more useful than the formal error provided by a one-dimensional inversion in that it would only be a measure for the sensitivity of the fitted atmospheric model to errors in the data. This, however, cannot be translated into a difference between a fitted parameter of the simplified atmosphere and it's true value at any given depth point. In the remainder of this paper we will therefore refrain from providing such error estimates.}

\section{Results}

\label{sec:results}
\begin{figure*}[htb]
\includegraphics[width=\textwidth]{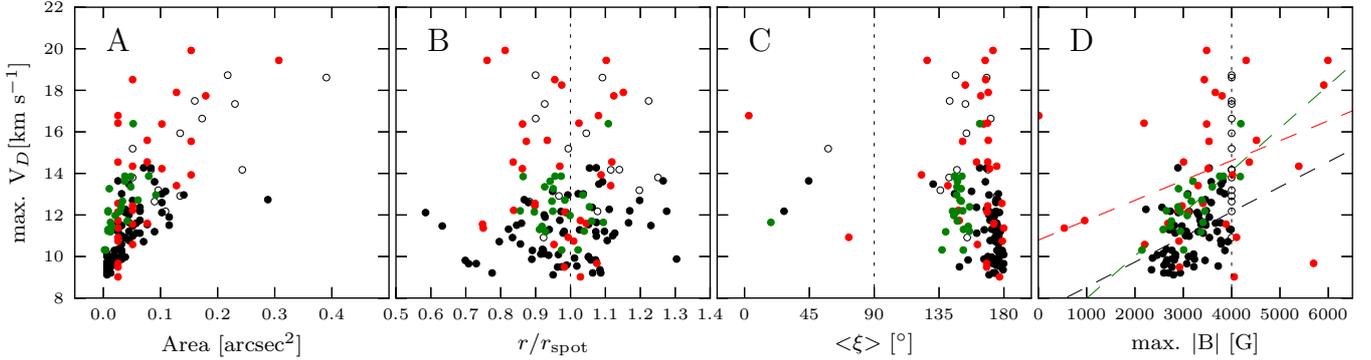}
\caption{From left to right: Maximum downflow velocity as a function of downflow area, distance to the center of the sunspot, inclination angle with respect to the sunspot polarity, and average magnetic field strength of the downflow patch for spot 1 (black), spot 2 (red), and the MHD-cube (green). Open black circles mark the downflow patches where the maximum field strength hits the upper limit of 4~kG set in the inversion of spot 1.}
\label{fig:downflows}
\end{figure*}

We now turn our attention to the inverted results. For spot 1 the whole spot was inverted at 0{\,\farcs}08$/$pixel, spot 2 was only inverted on a 0{\,\farcs}16$/$pixel grid. In addition, a \MURaM simulation of a sunspot by \cite{remp12} with a pixel resolution of 32~km in the horizontal direction and 16~km in the vertical direction was used to compare the results. To obtain maps that are directly comparable to the inverted Hinode SP maps, the optical depth scale of the \MURaM simulation was calculated using the line synthesis mode of the SPINOR code. The parameter values at the desired optical depth were then extracted to produce maps of the atmospheric quantities, comparable to those resulting from the inversions, but at the higher spatial resolution of the simulations.

Spot 1 was inverted first, using a fixed maximum Doppler velocity of 19\kms{} and a maximum magnetic field value $B$ of 4~kG. Several downflows, indicated by empty black circles in \fig{fig:downflows}, had to be excluded from the study because at least one parameter had reached it's limit. The limits were adjusted appropriately for the inversion of spot 2, so that no additional selection had to be made for that spot.

The magnetic polarity of spot 1 and the \MURaM simulation were both positive, that of spot two was negative. To allow for a comparison of the inversion results of all three, the magnetic field inclination $\gamma$ is expressed in a new coordinate $\xi=\gamma$ for the positive spots and $\xi=180-\gamma$ for the negative spot.

The most dominant feature of the inverted Doppler velocity maps, shown in \fig{fig:overview}, is the granular flow pattern that is clearly visible around both the sunspots. This contrasts with the penumbra, that shows many compact regions containing predominantly upflows in the inner and downflows in the outer penumbra, actually at the inner and outer ends of penumbral filaments \citep{tiwari13}. The downflows completely dominate over the upflows on the border between the penumbra and the granulation surrounding it, resulting in a clearly defined ring of downflows surrounding the entire sunspot.

In this ring of downflows, the average downflow speed is around 5\kms{}, but many compact regions are seen where the Doppler velocity exceeds 10\kms{}. The sound speed, related to the gas pressure $P_g$ and the mass density $\rho$ through
$$
C_s=\sqrt{\frac{\eta P_g}{\rho}},
$$
with $\eta=\frac{5}{3}$, was found to assume values of 5--8\kms, well below the observed Doppler velocity of these flows, indicating that these flows are supersonic. 

To study the properties of these supersonic downflows in more quantitative detail, contour maps at a level of 9\kms{} were made and all contours enclosing regions with a Doppler velocity exceeding 9\kms{} were considered to localize supersonic downflows and selected for further analysis. \corrone{The contours are based on bicubic spline interpolations of the discrete atmospheric values. Consequently, it is possible to create a contour which is smaller than a pixel (which is only 0.08" for spot 1 and 0.16" for spot 2). This does not imply that the downflow exists only in that single pixel, but it may mean that the imposed threshold of 9\kms{} was exceeded only in a single pixel within that downflow.}

\begin{figure*}[htb]
\includegraphics[width=\textwidth]{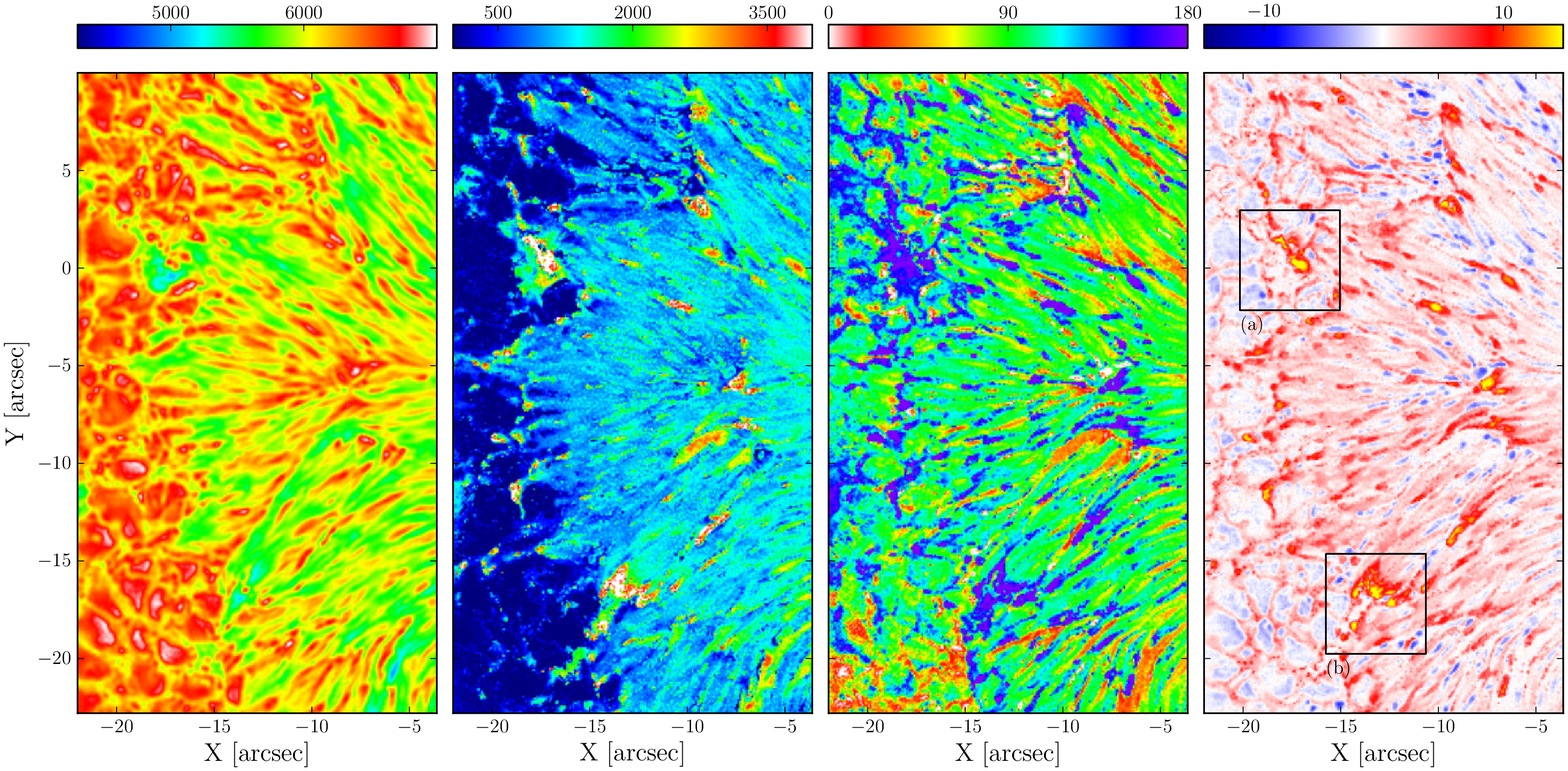}
\caption{Blow up of region 3 from \fig{fig:overview}. From left to right: temperature [K], magnetic field strength [G], inclination [$^\circ$], and line-of-sight velocity [\kms{}], all at $\tlc=0$. The boxes mark the downflow regions studied in \sref{flowstructure}.}
\label{fig:blowupB}
\end{figure*}

\subsection{Statistical Analysis}
A statistical study of the properties of the supersonic downflows, selected as described above, was carried out for all three spots. In \fig{fig:downflows} the downflow velocity is plotted as a function of a number of properties for spot 1 (black circles), spot 2 (red circles) and the \MURaM simulation (green circles). 

In panel A of \fig{fig:downflows}, the maximum downflow velocity is plotted as a function of the enclosed downflow area for all the selected supersonic downflows. Many of the downflows are less than 0.1~arcsec$^2$ in size, but a small number of much larger ones are also found, with areas of up to 1.5 square arcseconds and Doppler velocities of up to 22\kms. A dependence of the maximum downflow velocity on the patch size is clearly visible, for both the inverted spots and the \MURaM simulation, although the flow speeds found in the simulation are slightly lower and the downflow areas are smaller than the observed ones.

Panel B of \fig{fig:downflows} shows the maximum value of the Doppler velocity within the enclosed area for all supersonic downflows as a function of the distance of the downflow to the center of the sunspot in average sunspot radii $r_{spot}$, defined to be the radius of a circle that best fits the boundary between the penumbra and the granulation surrounding it. With the exception of a few, the downflows are all clustered around $r=r_{spot}$, coinciding with the outer boundary of the penumbra. Only a very small number of supersonic downflows are located inside or outside this ``ring''.

Panel C of \fig{fig:downflows} shows the maximum downflow velocity of each supersonic downflow as a function of the inclination angle $\xi$. According to the definition of $\xi$, an inclination angle of 0$^\circ$--90$^\circ$ corresponds to downflows with the same polarity as the umbra, whereas inclination angles between 90$^\circ$ and 180$^\circ$ denote opposite polarity downflows, for all spots. Clearly, the vast majority of the downflows harbor opposite polarity magnetic fields, with no dependence of significance on the downflow velocity. Supersonic downflows with the same polarity as the umbra are also observed in both the observed sunspots and the simulation, although their number is very small.

Panel D of \fig{fig:downflows} shows the maximum velocity in the downflows plotted as a function of the maximum magnetic field strength in the flow. A weak linear relation can be observed, that holds for both inversions and the \MURaM simulation, with correlation coefficients of 0.41, 0.37 and 0.76 for spot 1, spot 2 and the \MURaM simulation respectively. Remarkably high magnetic field strengths, in excess of 4~kG, are found in both the simulations and the inversions.

Only a very slight clustering in the azimuthal position was found for both the number of flows and the maximum flow velocity in both the observed spots (not shown). Since spot 1 was viewed at disc center and spot 2 at an angle of some 15$^\circ$, the clustering probably does not indicate a dependence of the flow properties on the viewing angle, but is more likely related to the symmetry properties of the spots.

\subsection{Supersonic Downflows}
\label{flowstructure}

To resolve the structure of the supersonic downflows optimally, in addition to spot 1, we also inverted a region of spot 2 (region (3) in \fig{fig:overview}), shown in \fig{fig:blowupB}, at an increased resolution of 0{\,\farcs}08$/$pixel. This region was found to contain some of the strongest and largest supersonic downflows including the two strongest downflows in all our data.

\begin{figure}[htb]
\includegraphics[width=\columnwidth]{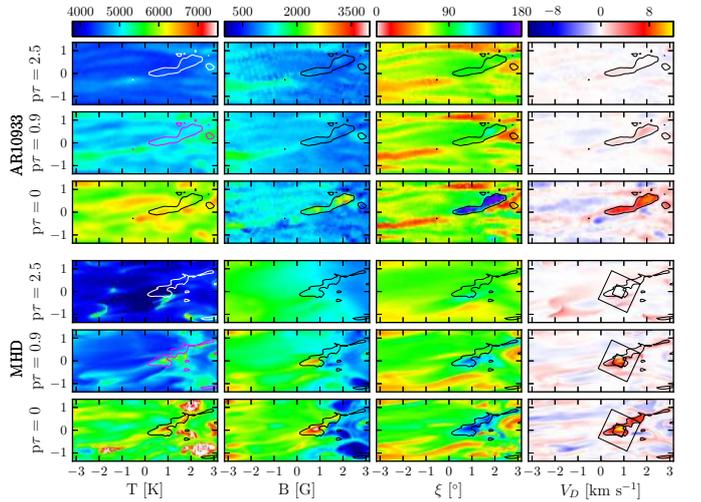}
\caption{Comparison between a downflow patch from the Hinode observations (top three rows, region (2) in \fig{fig:overview}) and the \MURaM simulations (bottom three rows). Maps of temperature (left column), magnetic field strength, inclination $\xi$, and line-of-sight velocity (right column) are displayed at $\tlc=[-2.5, -0.9, 0.0]$ (from top to bottom). The contour lines enclose the regions with line-of-sight velocities exceeding 5\kms{} in the deepest layer ($\tlc=0.0$). Tick marks are one arcsecond apart.}
\label{fig:ndf}
\end{figure}

The structure of the majority of the downflows appears to be fairly simple, with most of them harboring a magnetic field with a polarity opposite to that of the sunspot umbra. An example of such a downflow is shown in detail in the top three rows of \fig{fig:ndf} (region (2) in \fig{fig:overview}). The downflows can be easily identified by their yellow color at optical depth unity, indicating a flow velocity of $\ge$10\kms{}, and are readily confirmed to contain near vertical magnetic field of a polarity opposite to that of the umbra with magnetic field values of $\sim$3~kG. At optical depth unity, the flows appear to be slightly warmer than the material surrounding them.

All these distinctive features appear to be strongly height dependent, with the Doppler velocity decreased to sub-sonic values of only a few \kms{} already at $\tlc=-0.9$ and only a modest magnetic field enhancement remaining in the strongest downflows. At $\tlc=-2.5$, there is virtually no indication of the presence of most of the downflows in any of the inverted parameters.

Such downflows are found at the ends of all penumbral filaments, although in the case of simple filaments, with a single head and a single tail, they only reach speeds of 6 \kms{} on average, so that many of them remain subsonic \citep{tiwari13}. The supersonic downflows of this type are found at the ends of more complex filaments, particularly those with multiple heads that merge to form a single tail (or alternatively multiple filaments merge). Other than this they have qualitatively similar properties to the downflows at the ends of simple filaments. 

For comparison, a supersonic downflow from the \MURaM simulations is also shown in \fig{fig:ndf} (bottom three rows). Although in the upper atmospheric node the average temperature appears to be somewhat higher than what is retrieved from the simulations, the similarity between the simulated and the observed downflows is close, especially in the deepest atmospheric node. The \MURaM simulation presents us with the opportunity to follow the flow to deeper layers, allowing us to examine the structure below the visible surface. 

\begin{figure}[htb]
\includegraphics[width=\columnwidth]{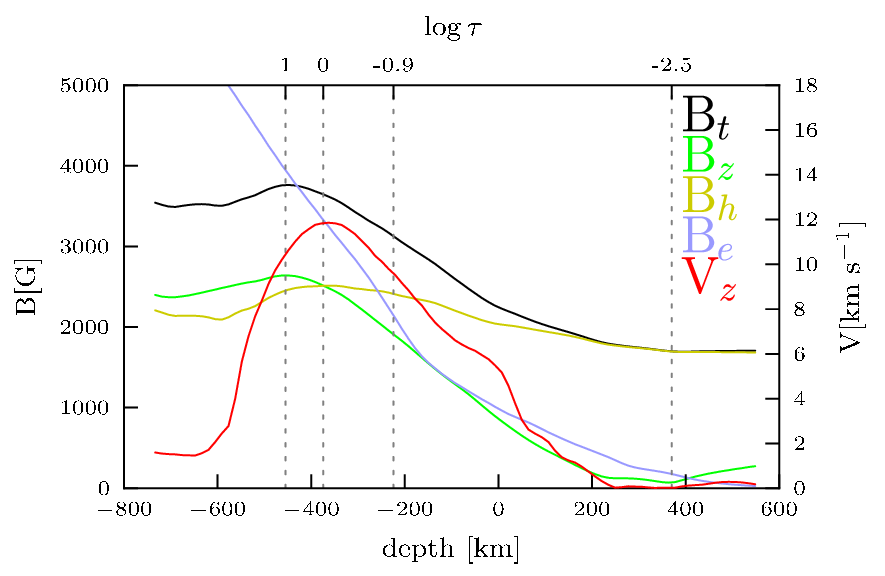}
\caption{Vertical structure of supersonic downflow from the \MURaM simulation. On the horizontal axis, the geometric height is indicated relative to the $\tau_c=1$ level of the atmosphere, horizontally averaged over the simulation box, indicating a Wilson depression in the downflow region of nearly 400 km.}
\label{fig:simdf}
\end{figure}

\fig{fig:simdf} depicts the vertical structure of the magnetic field strength, averaged over the downflow region indicated by the small box in the Doppler velocity map of the \MURaM simulated downflow in \fig{fig:ndf}. The magnetic field strength, $B_t$, increases with depth from a ``typical'' surface value of some 2~kG, to a maximum of 4~kG at an optical depth slightly larger than unity. From the location of the maximum value towards larger optical depth, the magnetic field strength first decreases slightly, then remains approximately constant. Interestingly, the change in the magnetic field strength appears to be almost completely accounted for by the changes in the vertical magnetic field strength $B_z$, whereas the horizontal magnetic field strength $B_h$ remains approximately constant throughout the atmosphere. 

The Doppler velocity $V_z$, co-plotted with the magnetic field in \fig{fig:simdf}, is seen to increase steadily from the top of the simulation box up to a maximum value of 12\kms{} at an optical depth just over 1, then it rapidly decreases to a much lower value of 1--2\kms{} and remains fairly constant towards deeper layers. The sudden drop in $V_z$ indicates the presence of a shock just below the visible surface.

The gas pressure $P_g$, expressed as the equivalent magnetic field $B_e=\sqrt{8\pi P_g}$, in the area immediately surrounding the downflow, indicated by the big box in the Doppler velocity map of the \MURaM simulated downflow in \fig{fig:ndf}, is co-plotted with the magnetic field in \fig{fig:simdf} for comparison. The value of $B_e$ equals that of $B_t$ just below the optical depth unity surface, at the maximum value of $B_t$ and close to the maximum of the Doppler velocity.

Only a few rapid downflows were found to contain magnetic field with the same polarity as the umbra. Panel C of \fig{fig:downflows} shows that one such downflow was present in the simulated spot and only two in spot 1 and spot 2 respectively. The low rate of occurrence may be due to the applied downflow selection threshold of 9\kms, as described in section \ref{sec:results}, much larger than the typical downflow speed of $\sim$1\kms{} of such ``anomalous'' downflows reported by \cite{kats10}.

In addition to the opposite polarity downflows, a number of very strong supersonic downflows that appear to have different properties are also found. These are discussed in the next section.

\subsection{Peculiar Supersonic Downflows}
\label{largeflowstructure}
The strongest and largest supersonic downflows in \fig{fig:blowupB} are found in regions (a) and (b). Like most other supersonic downflows they contain nearly vertical, strong magnetic field of opposite polarity. Different from most other supersonic downflows, however,  they contain also vertical magnetic field of the same polarity as the umbra. In addition, their structure has a significant vertical extent, making it visible in all height layers of the inverted atmosphere.

A common feature of such ``peculiar'' supersonic downflows is that they are all situated on the outer edge of the penumbra and are bordered on the outside (i.e. on the side facing away from the sunspot umbra) by an area of strong magnetic field of the same polarity as the downflow, such as a plage patch or a (micro)pore. This type of supersonic downflow is clearly less common than the ``regular'' ones, with only a few of them detected in spot 2, and none in spot 1 and the \MURaM simulation. We find them on both sides of spot 2, rendering an explanation in terms of projection effects unlikely.

\begin{figure}[htb]
\includegraphics[width=\columnwidth]{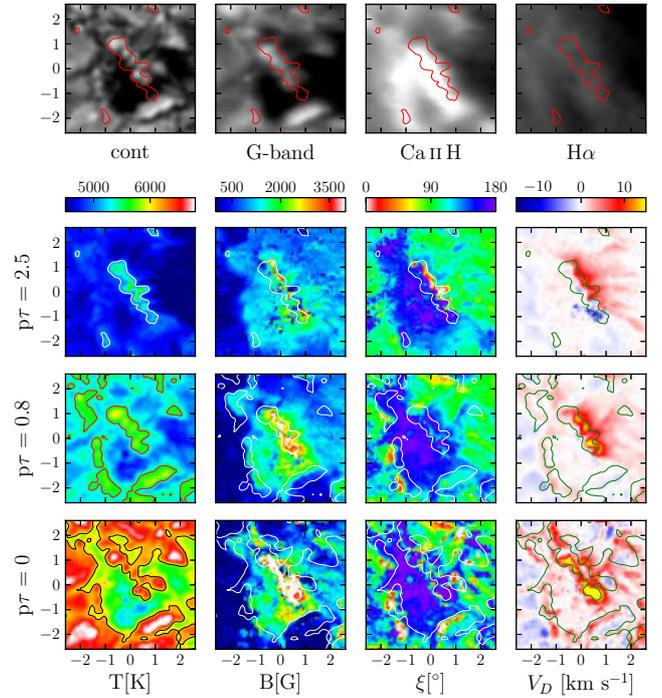}
\caption{Region from spot 2 (box (a) in \fig{fig:blowupB}). The top panel shows the continuum map of the inverted profiles (left), and the G-band,   \caiih{} images from Hinode BFI, and \halpha{} from Hinode NFI (right). The   bottom three rows show from left to right: temperature [K], magnetic field   strength [G], inclination [$^\circ$], and line-of-sight velocity [\kms{}], at different $-\tlc$ (i.e. p$\tau$) levels. Iso-temperature contour lines of 5000~K at p$\tau=2.5$ are overplotted in the $1^{st}$ and 2$^{nd}$ row, 5500~K at p$\tau=0.8$ in the 3$^{rd}$ row, and 6100~K at p$\tau=0$ in the bottom row. \corrtwo{The temporal evolution of the region surrounding the downflow in the G band, \caiih{} and \halpha{} (also shown in the top row) is shown in the attached movie.} }
\label{fig:df1}
\end{figure}

The highly structured character of the peculiar downflows is clearly visible in all physical parameters. \fig{fig:df1} shows a selection of inverted parameters and observations for one such downflow, situated in region (a) of \fig{fig:blowupB}. An elongated temperature increase of some 1000~K is clearly visible in the upper node of the inverted atmosphere. This temperature increase is less marked at greater optical depth and is virtually absent at optical depth unity, opposite to the behavior seen in ``normal'' downflows. 
The maximum magnetic field strength in the area of increased temperature is high, with a value of 3.5~kG in the upper node, increasing to \corrone{more than 7~kG} in the deepest node. Although most of the magnetic field found in the downflow area and in the external magnetic field patch has a polarity opposite to that of the sunspot umbra, the area showing the highest magnetic field strength is nearly vertically inclined and has the same polarity as the umbra. 

The umbral polarity field is organized in a ``sheet'', which is clearly visible in the upper atmospheric node, with the inclination angle $\xi$ reaching values close to 0$^\circ$ in some places, but is much less clearly marked in the deeper atmospheric nodes. The contours of constant temperature, drawn in \fig{fig:df1}, show that the sheet is not centered in the area of increased temperature, but is located more towards the sunspot, with the neutral line between the external magnetic field patch and the sheet accurately marking the long axis of the temperature contour.

Away from the downflow, in the direction of the sunspot, the magnetic field is found to be close to horizontal. In the downflow area, however, the inclination gradually increases towards the center of the downflow area, along with the Doppler velocity, peaking at 130$^\circ$ in the top node, and 180$^\circ$ in the lower nodes, in the same location as where the Doppler velocity reaches its maximum. From this location towards the center of the temperature contour, the inclination rapidly decreases to 0$^\circ$ while the field strength increases, forming the magnetic sheet. 

The line-of-sight velocity shows very similar behavior to the magnetic field strength, increasing from just below 10\kms{} in the upper node, to a maximum of 22\kms{} at optical depth unity. A small area of upflows of approximately $-$5\kms{} can also be seen in the upper node. The downflow velocity is at a maximum in an elongated area that is located on the contour of constant temperature and marks the location where the inclination is seen to peak before reversing direction.

With the exception of the azimuth, which appears to be very poorly determined, all quantities show a coherent, ribbon-like structure in the upper atmosphere, which appears to fragment significantly with depth. The only exception is the micro-turbulent velocity, which is small everywhere, except in the magnetic ribbon (not shown), where it increases rapidly as a function of optical depth, from values of $\sim$2\kms{} in the upper atmospheric node to approximately 10\kms{} at optical depth unity, while retaining a coherent structure at all heights. 

Close to the time the SP scanned across the downflow region, images recorded with the Broad-band Filter Imager (BFI) and the Narrow-band Filter Imager (NFI) in the G-band, the \caiih{} and the \halpha{} lines, show enhanced emission in the G-band and \caiih{}, but no brightening whatsoever in \halpha{}. In particular the brightening in the \caiih{} line closely resembles the shape of the inverted temperature rise. A time series of these data, available as online material, shows that this emission region is slowly but continually changing shape. It is continually present over a period of at least 30 minutes, indicating that the temperature increase and by inference the associated downflow are not the result of a transient phenomenon.

\begin{figure}[htb]
\includegraphics[width=\columnwidth]{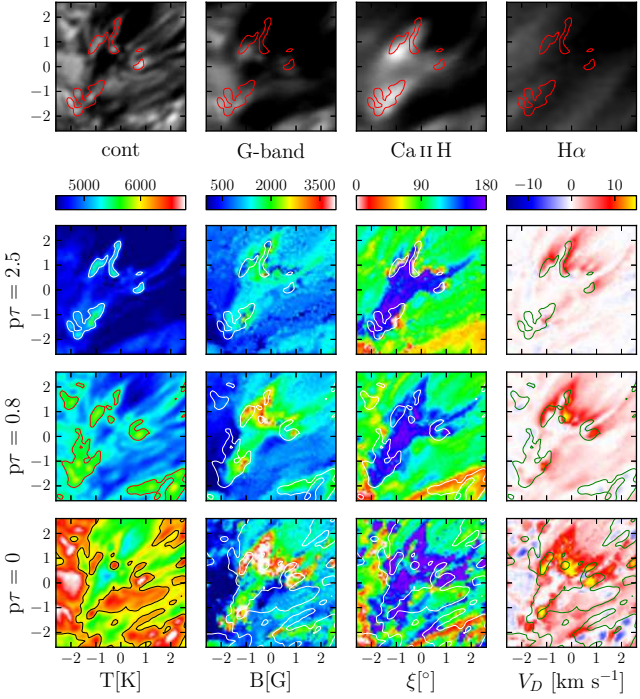}
\caption{Region from spot 2. Same as \fig{fig:df1}, but for box (b) in \fig{fig:blowupB}.}
\label{fig:df2}
\end{figure}
\fig{fig:df2} shows a second peculiar supersonic downflow, located in the lower half of region (2), which also shows clear signs of heating in the upper atmospheric node. This downflow is clearly more irregular and is distributed around the edge of a Y-shaped magnetic field patch.

The offset of the temperature increase with respect to the other parameters is somewhat smaller than for the downflow from \fig{fig:df1}, but it is nonetheless clearly visible, even though the regions showing the higher temperatures are more fragmented and compact. Here too, the magnetic field \corrone{strength} reaches very high values of \corrone{more than 7~kG} in the deepest layer and the downflow velocity a flow speed of 21\kms. The increase in the inclination before reversal of the field direction is also seen in this region at all heights.

Also in the second downflow, the morphology of the \caiih{} emission matches that of the increased temperature derived from the \FeI{} lines very well at the moment the Hinode-SP scanned the downflow region. The BFI time series in the \caiih{} line shows that the emission persists for at least 30 minutes, during which the bright fragments move around considerably, only matching the inverted temperature structure at the time of the scan, while no brightening is visible in \halpha{} in the NFI filtergrams.

The offset between the temperature increase and the area of maximum downflow was seen in four more supersonic downflows in spot 2, three of them located on the disk center side of the sunspot. Although these additional flows are clearly weaker and smaller than those in \fig{fig:df1} and \fig{fig:df2}, upon closer examination they all appear to show an arc of more inclined magnetic field in the middle atmospheric node. Two of them also show a polarity reversal of the magnetic field in the upper atmospheric node, as found in the strong downflows. Despite their less extreme appearance, their observational properties suggest that they contain a substructure similar to that of the two peculiar supersonic downflows discussed above, classifying them similarly as peculiar.

\section{Discussion}
\label{discussion}

The analysis of two real and one simulated sunspots confirms that strongly supersonic downflows are a common feature of sunspot penumbrae. Such downflows are located at or close to the outer penumbral boundary, generally where a number of penumbral filaments converge and probably harbor a portion of the Evershed return flow. Although they are supersonic, their association with penumbral filaments suggests that they may be the result of the merger of the tails of several penumbral filaments. They have properties similar to the generally weaker downflows found at the tails of all penumbral filaments. 

The dependence of the downflow velocity on the downflow area may be an indication that the downflows are not resolved in the observations and an average velocity is fitted by the inversion procedure. The slightly steeper relationship found in the data may well have to do with the higher spatial resolution of the simulations (which allows for smaller downflow areas), suggesting that a number of the observed rapid downflows are indeed not completely resolved. An alternative explanation is that since multiple penumbral filaments are joining up in one place to form a collective downflow, the more filaments join up, the larger and stronger that downflow becomes, resulting in the observed relation. The good agreement between the inversion results and the \MURaM sunspot simulations in panel A of \fig{fig:downflows} certainly speaks in favor of the latter explanation, although it is possible that the simulations themselves are not resolving the downflows. However, the data do not indicate that there is a specific size where the relation changes significantly. Since it is unlikely that the largest downflows, measuring 1.5~arcsec$^2$, are not resolved, we conclude that the relation is real.

The majority of the downflows do not extend much vertically and are not visible in the highest node of the inverted atmosphere. This behavior is markedly different in the simulation, where almost all downflows show a significant vertical extent and a temperature increase of approximately 1000~K at $\tlc=-2.5$. The cause may well be in the approximations made in the treatment of the energy equation in the \MURaM simulation, indicating that more detailed simulations may be required to reproduce the behavior that is actually observed.

Most of the supersonic downflows are only supersonic at the location of the deepest node in the inverted atmosphere. At optical depth unity they show an increased temperature compared to the surrounding atmosphere and contain strong magnetic field, with a field strength occasionally exceeding 4~kG, with an orientation close to opposite that of the magnetic field in the sunspot umbra. This behavior is seen in the inverted downflows as well as the \MURaM simulated ones, suggesting that such high field strengths do exist in the outer penumbra at locations of supersonic downflows.

Since the flow is downwards and supersonic, deceleration takes place in the form of a shock. The vertical structure of the \MURaM simulation indicates that the shock, marked by a sharp drop in the vertical velocity in \fig{fig:simdf}, is found in most downflows at an optical depth between 10 and 100. The depth of the shock below the surface depends on the speed of the downflow \corrone{and coincides with the depth at which the gas pressure of the atmosphere surrounding the downflow becomes similar to the magnetic pressure in the downflow. Below the shock, the field strength remains approximately constant.}

\corrone{The strong magnetic fields are probably the result of intensification of magnetic field by the collapse of magnetized flux concentrations \citep[e.g.][]{parker1978}. Although in the quiet Sun field strengths of only 1-1.5 kG are produced by convective downdrafts, in penumbral downflows most magnetic elements carry a highly supersonic downflow of 10-15\kms{}, resulting in significantly higher magnetic field strengths. In addition, the correspondingly low gas pressure in the magnetic elements, lowers the optical depth unity surface, exposing deeper, stronger fields. This mechanism offers a natural explanation for the relation between the Doppler velocity of the flow and the magnetic field strength seen in \fig{fig:downflows}}. 

Despite the presence of a shock just below the visible surface and a significant Wilson depression of some 400 km, no significant temperature rise is observed in the downflows compared to the material surrounding them. Since the configuration of a downflow is not unlike that of a typical magnetic flux tube, a temperature rise caused by radiation flowing in horizontally through the walls of the Wilson depression is expected \citep{spruit1976}. A possible explanation for the absence of such a temperature rise is that the material is moving through the downflow region so fast that the radiation does not have enough time to heat the material significantly before it flows through the optical depth unity surface.

As can be seen in \fig{fig:simdf}, the horizontal magnetic field strength, $B_h$, remains approximately constant throughout the atmosphere. This phenomenon is the consequence of the continuity of the current in the downflow, which prevents any changes to the horizontal magnetic field strength \citep{1979cmft.book.....P}. The horizontal and vertical magnetic field structures of the downflow are therefore effectively decoupled, resulting in a semi one-dimensional situation. Although significant variations are found in $B_x$ and $B_y$, probably caused by a large amount of twist, the total horizontal component of the magnetic field is surprisingly constant and variations in the total field strength $B_t$ are almost entirely caused by changes in the strength of the vertical field component $B_z$. 

The maximum value of the magnetic field is set by the \corrone{gas pressure} of the atmosphere around the downflow at the height of the shock and \corrone{thus} is dependent on the depth of the shock. Using the linear relations between downflow speed and field strength in \fig{fig:downflows}, it is possible to extrapolate the field strengths observed in the downflows in the \MURaM simulation to the vertical velocities observed in spot 1 and spot 2. From such an extrapolation, a magnetic field value of 6-7~kG is expected, in good agreement with the values actually found from the inversions. The very significant horizontal component, clearly visible in the \MURaM simulation, is not present in the inversion results. The absence of the horizontal field component may be because the magnetic field is not significantly twisted, but is more likely caused by signal cancellation as a consequence of insufficient spatial resolution.

\subsection{Peculiar downflows}
There are a number of ``peculiar'' downflows found bordering external patches of strong magnetic field of the opposite polarity from the umbra. Unlike the ``regular'' supersonic downflows, the peculiar supersonic downflows are seen at all heights in the inverted atmosphere and are accompanied by a temperature increase in the highest atmospheric node of around 1000~K. This heating is seen in the \FeI{} lines at 6301.5 and 6302.5~\AA\ and in the \caiih{} line at 3964~\AA. A time series of filtergrams in the \caiih{} line indicates the heating is long lived and only slowly varying. The absence of excess emission in the \halpha{} line suggests that the heating is confined to the upper photosphere and probably does not extend much into the chromosphere. These peculiar downflows are found to contain very strong magnetic fields with field strengths exceeding 7~kG. Since this type of downflow is not present in the \MURaM simulations, a direct comparison with magneto-hydrodynamic models could not be made.

\begin{figure}[htb]
\includegraphics[width=\columnwidth]{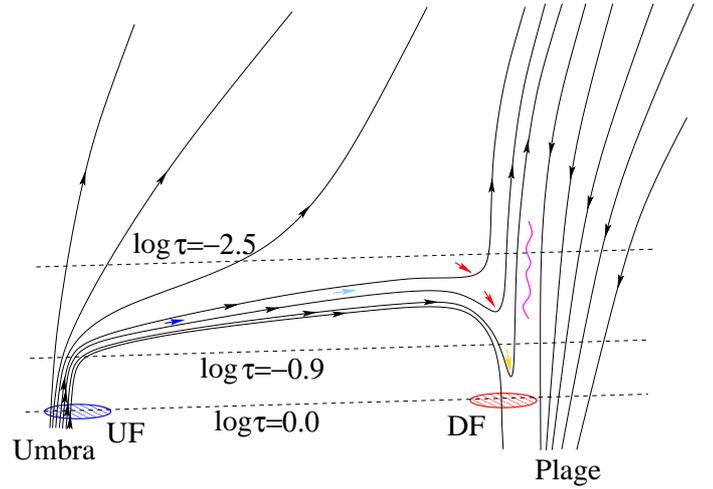}
\caption{Possible magnetic field configuration of peculiar supersonic downflow. \corrone{The black lines indicate the magnetic field, the black arrows the magnetic field direction. The gas flow is indicated with yellow, red and blue arrows, where blue indicates a blue-shift (upflow) and red and yellow a red-shift (downflow) when viewed from above, similar to the color coding used in Figure~\ref{fig:df1}. The magenta wavy line indicates the possible location of a current sheet, responsible for the heating in the upper layer of the atmosphere.}}
\label{fig:cartoon}
\end{figure}

The proposed scenario giving rise to these peculiar downflows is depicted in a sketch in \fig{fig:cartoon}. An external patch of magnetic field provides a magnetic barrier that the material flowing in the magnetized channels of the penumbral filaments, carrying the Evershed flow, is not able to cross. The flow in the magnetized Evershed channel is not perfectly aligned with the magnetic field, resulting in the advection of umbral magnetic field towards the tail of the penumbral filament. The advection of magnetic field leads to an increase in the magnetic field strength, until the field strength is sufficient to affect the flow and an equilibrium is reached.

In the majority of the penumbral filament tails, the field is dragged under by the flow, resulting in the expansion of the magnetic field behind the penumbral filament. In the above situation, however, the magnetic field is unable to expand due to the magnetic barrier and instead piles up in a layer of intensified, ``umbral'' polarity field between the downflow and the magnetic barrier field.

The inclination of the advected field is easily seen to be compatible with the observed behavior, with a slight and gradual initial decrease, followed by a steep increase to a nearly vertical orientation, pressed tightly against the external magnetic field patch.

The formation of strong magnetic fields with opposite polarity in such close proximity of each other will induce the formation of a current sheet between them, illustrated with a pink wavy line in the sketch. Finite resistivity will ultimately cause this current sheet to dissipate, resulting in the observed heating. 

Two scenarios are proposed for this dissipation: In the slow scenario, the current sheet simply dissipates by Ohmic resistivity and heats the area immediately bordering the neutral line. There is no acceleration associated with this process and the observed acceleration of the downflows is driven purely by redirection of the flow and gravity. 

In the fast scenario, the field may reconnect in the upper photosphere, causing the field configuration to change and the material to be accelerated by the resulting magnetic tension forces. This mechanism should produce both upward and downward acceleration, resulting in an up- as well as a downflow. 

In both cases, the heating is sustained by advection of new field to the tail of the filament. The observed heating, the offset of the observed downflow with respect to the location of the heating and the observed increase in the magnetic field inclination followed by a sharp drop are all compatible with the configuration outlined above.

For reconnection taking place in the upper photosphere, the Hinode/SP lines would sample mainly the downflow. Such a rapid, reconnection-driven downflow might help to explain why such anomalous downflows are among the fastest observed. 

Although the increased inclination of the magnetic sheet at $\tlc=-0.8$ suggests that the opposite polarity field may no longer be present at this height, possibly because it is dissipated by magnetic reconnection, it is also possible that the structure is spatially unresolved. Some of the upflow expected from magnetic reconnection is observed, but it is only present in a small fraction of the area of increased temperature. The contribution of magnetic reconnection to the heating and the downflow, although likely, cannot be conclusively established or dismissed with the available information. 

\corrone{The presence of substructure is also suggested by the behavior of the microturbulent velocity. Although the supersonic microturbulent velocities recovered may be real, it is more likely that the broadening of the line profiles that is responsible for their appearance is the product of (partial) signal cancellation of the Stokes Q, U and V profiles due to lack of resolution, reducing the polarization signals, but not the line broadening. Due to the large depth of formation of the wings of the line profile, the Zeeman splitting that may be present is obscured by higher layers, so that the broadening can no longer be explained with a strong magnetic field and can only be interpreted as microturbulence instead. In this scenario, the recovered magnetic field values may underestimate the true field strength.}

\subsection{Strong magnetic fields}
\label{sec:strongmag}
The very high magnetic field values seen in the deep layers of many of the supersonic downflows in the inverted atmospheres, are significantly stronger than even the strongest solar magnetic fields observed so far, putting the reliability of the inversion results into question. \cite{2006SoPh..239...41L}, who searched historical measurements from Mt Wilson and Rome observatories covering many decades for spots with particularly strong fields, found that only 0.2\%\ of all spots have field strengths larger than 4~kG, with the strongest umbral field ever measured being 6.1~kG. The values in excess of 7~kG measured in the downflows would therefore make these candidates for the strongest fields ever measured on the Sun. For such strong magnetic fields, it is important to consider what the effect of the assumptions and approximations made by the inversion code might be on the validity of the result. 

In the absence of time dependence, the assumed approximation of hydrostatic equilibrium, needed to calculate the opacity, is the only practical one. In the presence of very strong magnetic fields, however, gas pressure and gravity are probably not the dominant contributors to the momentum equation. The error in the derived density stratification caused by this is likely to significantly affect the formation of the line and therefore the inverted atmospheric stratification. However, many downflows containing high field strengths are also found in the \MURaM simulation, where no such problem exists.

The very high values of the magnetic field strength are predominantly based on the very broad wings of the Stokes $V$ profiles\corrone{, such as the profiles shown in Fig~\ref{fig:profiles1}}, that can only be produced by a strong magnetic field near optical depth unity. An explanation in terms of a magnetized, turbulent flow is also possible, but requires a highly structured magnetic field, to avoid the cancellation of signal.

Alternatively, the Zeeman splitting that appears to be visible in some of the Stokes $I$ line profiles could be the result of multiple atmospheric components with large differences in the Doppler velocity. It is, however, not clear that such a model would be able to reproduce both the 6301.5~\AA\ and the 6302.5~\AA\ line profiles, with their different Land\'e factors, simultaneously for all Stokes parameters with the observed accuracy, although no attempt to produce such a model was made.

\begin{figure}[htb]
\includegraphics[width=\columnwidth]{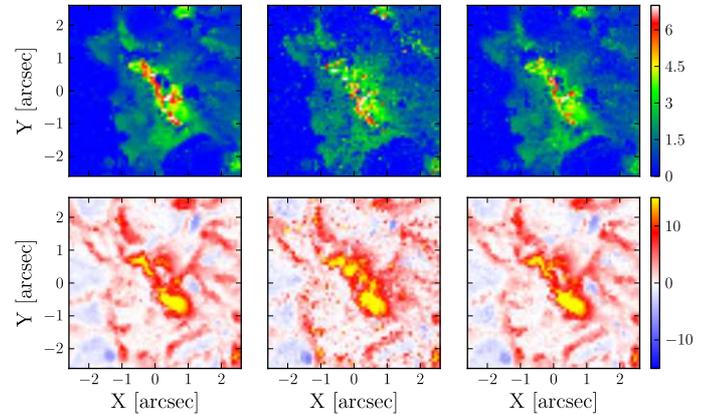}
\caption{Inverted magnetic field strength (top row) and line of sight velocity (bottom row) for the original inversion (left), test case 1 (middle) and test case 2 (right).}
\label{fig:robust}
\end{figure}

\corrone{To gain more confidence in the results, a test of the robustness of the solution to noise was carried out for a small region containing downflow 1. This was done by creating two new datasets: one from the original data by increasing the noise by a factor of 2.25, the second by using the fitted convolved profiles from the best fit to the original data and adding a similar amount of noise to it as is present in the original data. Both data sets were then inverted in the same way and compared. The data in the first test has a higher noise level and thus the inversion result should contain larger errors than the original inversion, the data in the second test was generated from fitted profiles, which can be reproduced exactly by the inversion code, so that the difference between the inversion result and the original inversion result will be smaller than the error in both results. A comparison of the magnetic field strength at optical depth unity is shown in Figure~\ref{fig:robust}. Although clearly there are differences, the results are very similar in both amplitude and appearance, suggesting that the sensitivity of the results presented in this paper to noise is not very high.}

The relation between the maximum downflow velocity and the maximum magnetic field strength shown in panel D of \fig{fig:downflows} suggests that the high magnetic field values are compatible with field intensification in the deep photosphere as found in the \MURaM simulation, and cannot easily be dismissed as merely the result of errors. Although the value of the magnetic field strength may well contain a significant error, it is nonetheless likely that the very high values returned by the inversion code are representative of the actual field strength in very fast downflows.

\section{Conclusions}
We have analyzed two observed and a simulated sunspot using Hinode SP scans and a \MURaM simulation. To recover the maximum amount of information from the observations, we interpolated the data to a finer grid before inverting them using the spatially coupled version of the SPINOR inversion code.

We find that this new approach to the inversions significantly improves the consistency of the results and increases the high-frequency information in most of the inverted atmospheric parameters. 

The inversion results indicate that sunspot penumbrae are surrounded by a ring of compact regions harboring supersonic downflows. The magnitude of the velocities found in these areas is in reasonable agreement with the maximum flow speeds of the horizontal component of the Evershed flow found in earlier studies, with values of 10--15\kms. The magnetic field in these downflows is unusually strong, with inverted field strengths typically between 2.5 and 4.5~kG, in good agreement with those found in the \MURaM sunspot simulation.

In areas where an extended region of strong magnetic field is present just outside the penumbra, the flow velocity is substantially higher than in the majority of the downflows, with peak values of 22\kms{}. The combination of heating in the upper atmosphere, unusually strong magnetic field values in excess of $7$~kG and the presence of small-scale opposite polarity field appears to indicate the presence of a current sheet, possibly induced by the accumulation and intensification of penumbral magnetic field by the Evershed flow. These current sheets are found to be common and are sustained over significant periods of time.

\begin{acknowledgements}  We thank M. Rempel for providing the sunspot simulation used in this paper and R. Cameron for many helpful discussions on the interpretation of the \MURaM simulation. This work has been partially supported by the WCU grant number R31-10016 funded by the Korean Ministry of Education, Science and Technology. Hinode is a Japanese mission developed and launched by ISAS/JAXA, collaborating with NAOJ as a domestic partner, NASA and STFC (UK) as international partners. Scientific operation of the Hinode mission is conducted by the Hinode science team organized at ISAS/JAXA. This team mainly consists of scientists from institutes in the partner countries. Support for the post-launch operation is provided by JAXA and NAOJ (Japan), STFC (U.K.), NASA, ESA, and NSC (Norway). 
\end{acknowledgements}

\bibliography{ms}

\end{document}